\documentclass{PoS}

\title{Latest results from the Pierre Auger Observatory}

\ShortTitle{Latest results from the Auger Observatory}

\author{\speaker{Esteban Roulet}, for the Pierre Auger Collaboration
\thanks{Full author list available in
{ http://www.auger.org/archive/authors\_2010\_11.html}.
Work partially supported by ANPCyT PICT 1334 and by CONICET PIP 01830.
Dedicated to the memory of H\'ector Rubinstein and his passion for
the physics of astroparticles.}\\
        CONICET, Centro At\'omico Bariloche\\
Bustillo 9500, Bariloche, 8400, Argentina\\
        E-mail: \email{roulet@cab.cnea.gov.ar}}

\abstract{Recent results obtained with the Pierre Auger Observatory are
  described. These include measurements of the spectrum, anisotropies and
  composition of ultra-high energy cosmic rays. The ankle of the spectrum is
  measured at $4\times 10^{18}$~eV and a suppression above $3\times
  10^{19}$~eV consistent with the GZK effect is observed.  At energies above
  $5.5\times 10^{19}$~eV a correlation with the distribution of nearby
  extragalactic objects is found, including an excess around the direction of
  Centaurus~A, the nearest radio loud active galaxy.  Measurements of the
  depth of shower maximum and its fluctuations suggest a gradual change
  in the average mass of the primary cosmic rays (under standard
  extrapolations of hadronic interaction models), being the results consistent
  with a light composition consisting mostly of 
protons at few$\times 10^{18}$~eV and approaching
  the expectations from iron nuclei at $4\times 10^{19}$~eV. Upper bounds on the
  photon fraction and the neutrino fluxes are also obtained.
}

\FullConference{Quarks, Strings and the Cosmos - H\'ector Rubinstein Memorial Symposium\\
		August 09-11, 2010\\
		AlbaNova, Stockholm, Sweden}

\begin{document}

\section{Introduction}

The Pierre Auger Observatory, built near the town of Malarg\"ue in Argentina,
has been gathering data since January 2004 \cite{auger}. 
It reached its baseline design
covering 3000~km$^2$ with 1600 water Cherenkov detectors overlooked by 24
fluorescence telescopes by mid 2008 and by the end of 2009 had accumulated a
total exposure of about twenty thousand km$^2$~sr~yr, much larger than that of
all previous air shower experiments combined. 
 The surface detector has a duty cycle of almost 100~\%,
 collecting then the vast majority of the data which are used  for spectrum
 measurements and anisotropy searches.
On the other hand, 
simultaneous observations with both the fluorescence and surface detectors are
possible for $\sim 15$\% of the events (those observed during moonless  nights
with no clouds), for which the longitudinal development in the atmosphere
as well as  the lateral profile on the ground can be measured. 
This allows the cross
calibration between the two detection techniques, since the UV fluorescent
light emitted by the nitrogen molecules excited by the electromagnetic
component of the air shower provide an almost calorimetric measurement of the
 energy of the primaries. It also allows  to determine the depth of maximum
 development  of the shower, which encodes precious information on the
 composition of the primaries and the properties  of the first hadronic
 interactions.  The studies of the cosmic rays at the
 highest energies with the Auger Observatory has already allowed to start
 addressing many of the old questions that motivated its construction by
 measuring the features present in the spectrum, searching for anisotropies in
 the cosmic ray arrival directions distribution or constraining the
 composition of the primary cosmic  rays.

\section{Spectrum}

\begin{figure}
\centerline{\includegraphics[angle=0,width=0.55\linewidth]{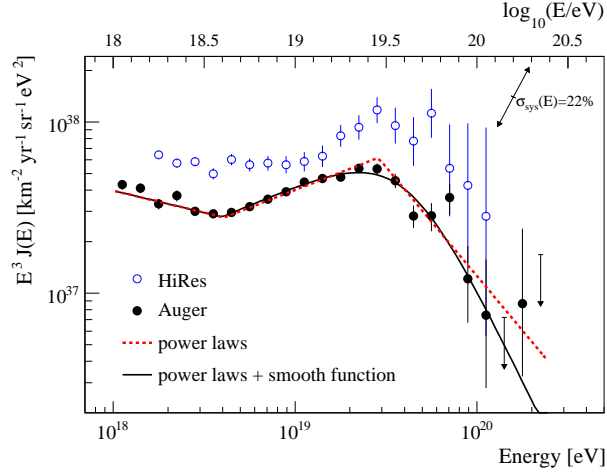}}
\caption{Spectrum measured above 1~EeV (solid dots)  and power law fits
with breaks  (dotted line). For comparison also the HiRes measurements are
displayed (open dots).}
\label{spectrum}
\end{figure}

In order to determine the cosmic ray spectrum, a reliable estimate of the
exposure is necessary, and hence a strict event selection is performed
requiring that the detector with the largest signal be surrounded by a full
hexagon of working detectors (for high energy anisotropy searches instead, a
relaxed trigger requires only 5 active detectors around the `hottest' one and
that the shower core be contained in an active triangle). 
 Events with zenith angles below $60^\circ$ are
used in the following studies and in this case the surface array is fully
efficient only above 3~EeV (where 1~${\rm EeV}\equiv 10^{18}$~eV), in which case all
showers trigger at least three detectors and can hence be reconstructed. Below
this energy the surface detector 
efficiency becomes less certain (depending in particular on the 
 composition of the cosmic rays), so that the 
spectrum is obtained instead using  hybrid events. 
The resulting measured spectrum \cite{augerspec} 
above 1~EeV is shown in Fig.~\ref{spectrum}. 
A piece wise fit using power laws (d$N/$d$E\propto E^{-\alpha}$) shows that
there are two clear transitions at 4.1~EeV and 29~EeV. The first feature 
is the so-called ankle, in which the spectral index changes from
$\alpha=3.26\pm 0.04$ to $2.59\pm 0.02$, and the second feature involves a
transition to a much steeper spectrum (the power law fit leading to
$\alpha=4.3\pm 0.2$), with the spectrum falling to half the value
that would be obtained from an extrapolation of the lower energy fit  at
$E_{1/2}\simeq 40$~EeV. 
One has to keep in mind that systematic effects on the
energy determination amount to 22\%, and are hence significant. In particular,
the different normalization of the spectrum measured by the HiRes experiment,
also shown for comparison in fig.~\ref{spectrum}, is most likely due to a
systematic energy mismatch between the two experiments.

The physical origin of the ankle is still uncertain, being the main candidate
scenarios to explain this feature 
 those relating it to the transition from a dying
galactic component to a harder extragalactic component becoming dominant, or
alternatively the so-called dip-scenario \cite{dip}, 
in which cosmic rays are assumed to be
extragalactic protons down to energies below 1~EeV and the concave shape
observed arises from the effect of energy losses by pair creation with cosmic
microwave background (CMB)
photons. To properly fit the observed spectral shape this last scenario
requires soft spectra at the sources ($\alpha$ at the source being typically
$\alpha_g\simeq 2.4$-2.7) and/or strong evolution of the sources with
redshift, what makes the distant sources intrinsically brighter (or more
abundant) so that a larger fraction of the observed protons come from far away
and are hence more affected by interactions with CMB photons. Also an upward
shift of the energy scale of Auger by $\sim 40$\% would be required in this
scenario to fit the location of the dip in the spectrum.

Anisotropy  measurements will help to distinguish among the two scenarios,
because the galactic/extragalactic transition  may lead to measurable dipole
type patterns in the arrival direction distribution resulting from the diffusive
escape of the galactic cosmic rays, and already significant constraints have
been obtained by Auger at EeV energies \cite{bonino}.
 Also composition measurements are
important because galactic cosmic rays at EeV energies are expected to be
dominated by heavy nuclei, since their confinement by galactic magnetic fields
is a rigidity dependent effect.   Enhancements of the Auger observatory to
improve the sensitivity down to energies of $10^{17}$~eV, such as the HEAT
fluorescence detectors or the AMIGA infill and muon detectors,  will help to
shed light on these issues in the near future. 

The second feature mentioned, i.e. the suppression observed at the highest
energies, is similar to the expectations from the so called GZK effect 
associated to the attenuation of extragalactic protons by photo-pion production
off CMB photons  (this suppression was 
predicted by Greisen and Zatsepin and Kuz'min just after the
discovery of the CMB). Also the 
 photodisintegration of nuclei as heavy as iron would lead to similar features,
while lighter
nuclei would instead show a suppression down to a lower energy threshold, in
 approximate proportion to their masses. Note that a
 change in the injection spectrum at the sources may also
contribute in part to the observed spectral shape.

\section{Anisotropies}

Searches for localized anisotropies are motivated by the fact that
cosmic ray trajectories in galactic and extragalactic magnetic fields
become straighter as the energy increases, being for instance the
typical deflection for a nucleus of charge $Z$ traveling a distance
$L$ in the galactic field
(for which the regular component has a strength of $\sim 3\ \mu$G
coherent over scales $\sim {\rm kpc}$) of
\begin{equation}
\delta\simeq 3^\circ Z\left(\frac{B}{3\ \mu{\rm
G}}\right)\left(\frac{L}{\rm kpc}\right)\left(\frac{60\ {\rm
EeV}}{E}\right).
\end{equation}
This gives then the hope that cosmic ray astronomy may become feasible
 at ultra-high energies. On the other hand, since above the GZK
threshold the energies of extragalactic cosmic rays are significantly
attenuated as they propagate through the cosmic photon backgrounds,
setting a sufficiently high energy threshold implies that only sources
within a relatively close-by neighborhood can contribute to the fluxes
observed at Earth. For instance, for a uniform distribution of proton
sources  90\% of the cosmic rays reaching the Earth with  energies
above 60~EeV should have been produced within about 200~Mpc \cite{horizon},
and comparable `GZK horizons' are also found in the case of Fe
sources. Then, an efficient way to search for an anisotropy pattern, before any
individual source clearly stands up above the background, is to
look for a correlation within a certain angular window between the
arrival directions of the  events above a certain
threshold energy and the location of a certain type of candidate 
sources within a given distance.

\begin{figure}
\centerline{\includegraphics[angle=0,width=0.65\linewidth]{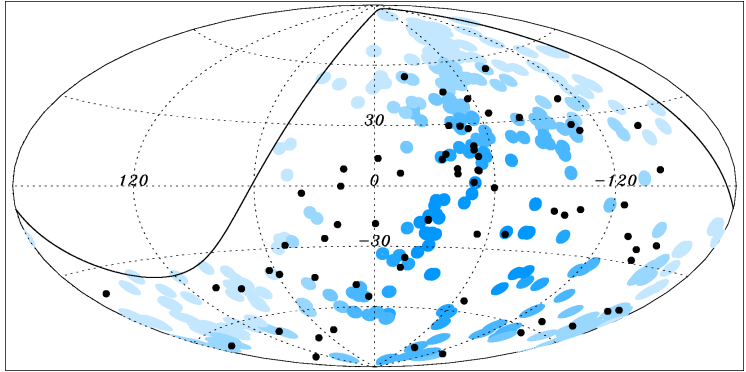}}
\caption{Arrival directions of the events above 55~EeV (dots) and $3.1^\circ$
  circles around the directions towards AGN in the VCV catalog closer than
  75~Mpc. }
\label{vcv}
\end{figure}

One class of potential
source population that may be able to accelerate particles up to these extreme 
energies is the Active Galactic Nuclei (AGN), consisting of the
supermassive black holes (with masses up to $\sim 10^9M_\odot$)
accreting matter in the center of galaxies and emitting powerful
jets. An analysis performed by the Auger Collaboration
\cite{science,astropart} indeed established a
 correlation with the AGN in the V\'eron Cetty and V\'eron (VCV) catalog (which
 is actually a compilation of catalogs). This correlation was most significant
 for events above 55~EeV and angular separations of less than 
$3.1^\circ$ from AGN
 closer than 75~Mpc. In the latest study with data up to the end of
 2009, the fraction of events
 correlating within those parameters (excluding the events from the initial
 period used to fix those values) is $38^{+7}_{-6}$\%, well above
 the 21\% that would be expected if the distribution were  isotropic
 \cite{agnnew}. A map of the observed arrival directions (dots) in galactic
 coordinates is shown in fig.~\ref{vcv}, displaying also circles of
 $3.1^\circ$ radius  around the location of nearby VCV AGN. One finds that 29
 out of the 69 events do indeed fall inside one of the circles. 
 Note that due to  obscuration effects the
 catalogs are particularly incomplete near the galactic plane, and
 hence it is understandable that most of the events within 
 $10^\circ$ of the galactic plane do not correlate with objects in
 the catalog.

\begin{figure}
\centerline{\includegraphics[angle=0,width=0.55\linewidth]{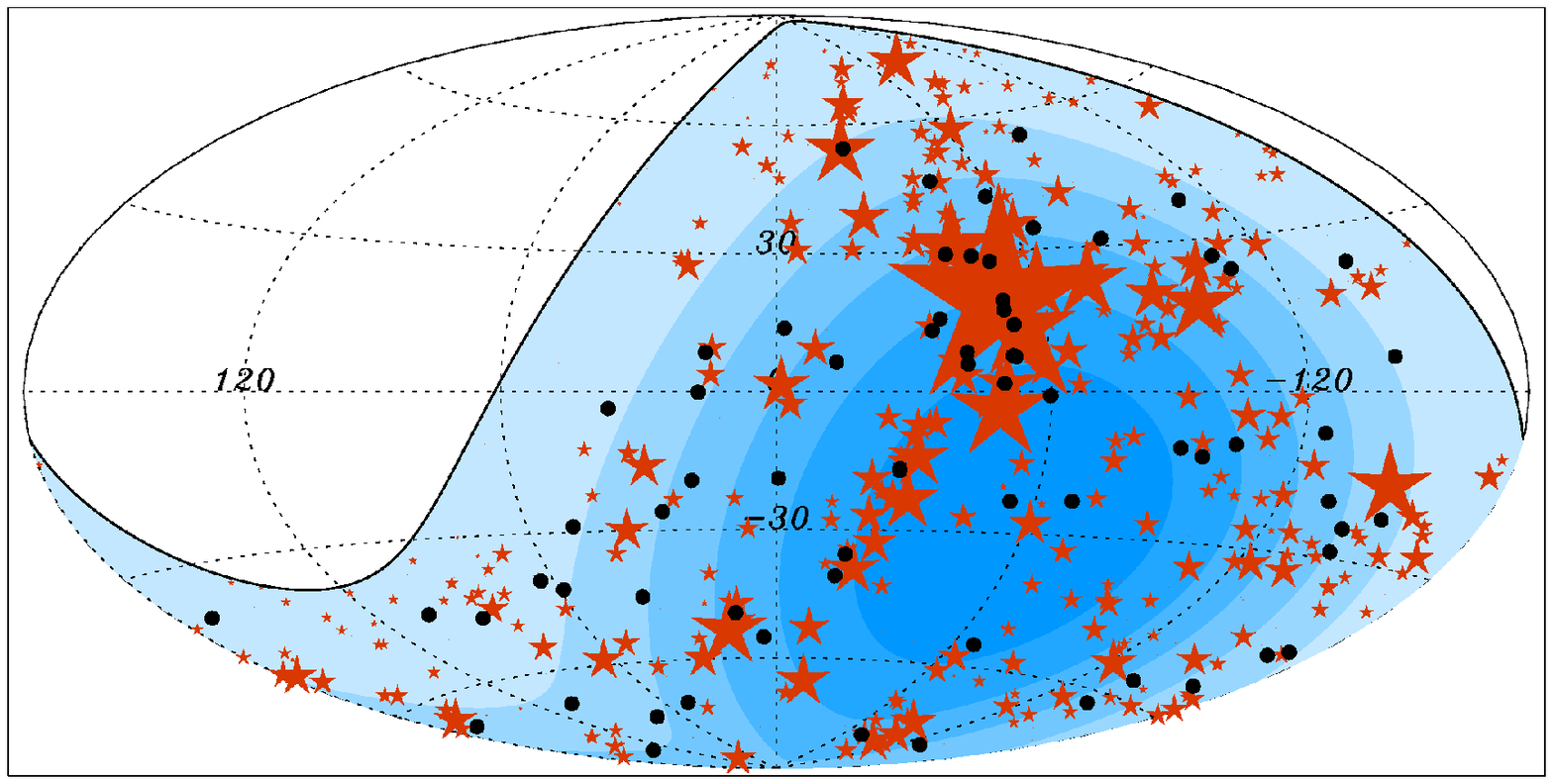}\ \ \ \  
\includegraphics[angle=0,width=0.28\linewidth]{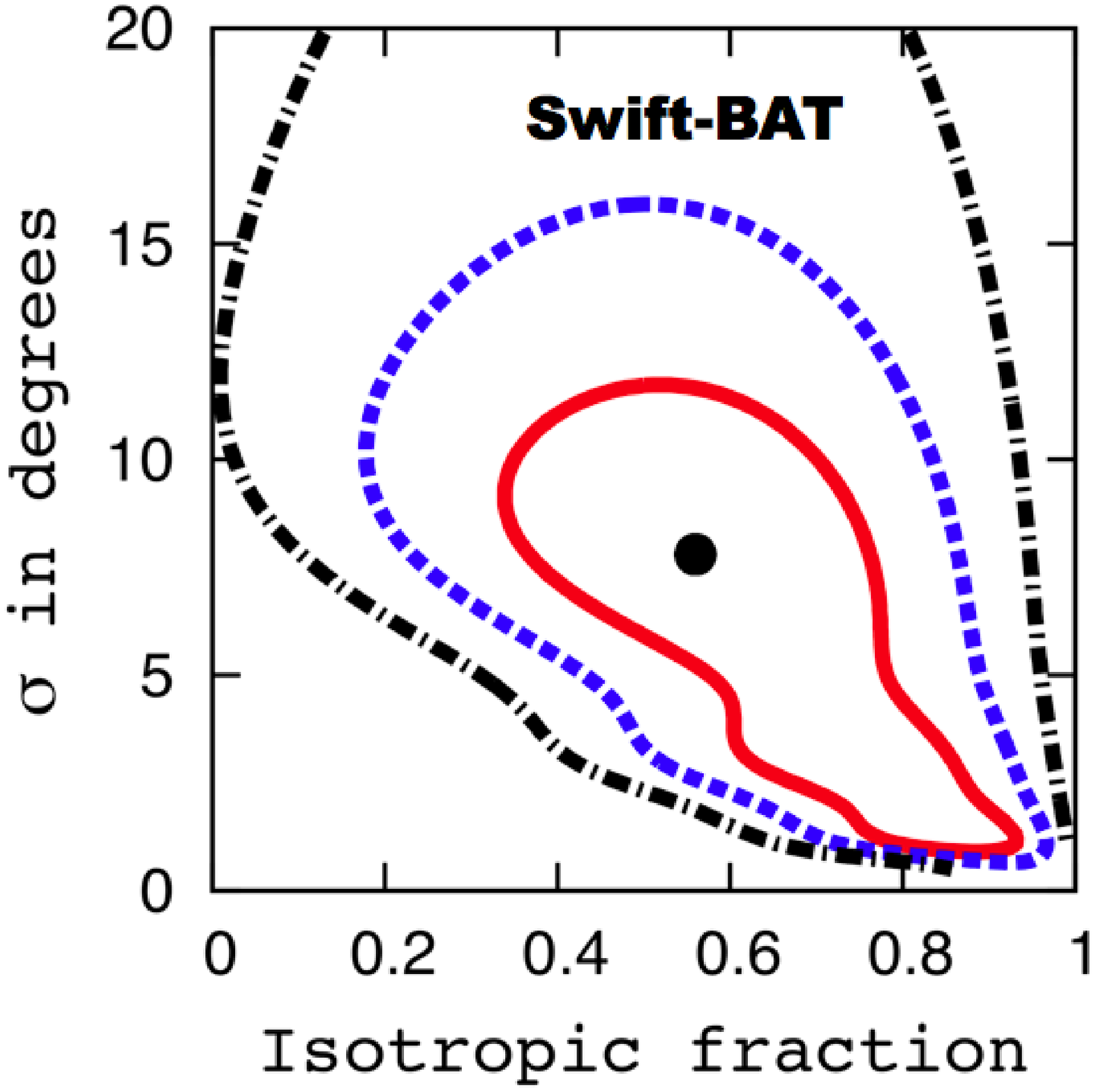}}
\caption{Left panel: map of arrival directions of the events above 55~EeV and
  AGNs observed in X-rays by SWIFT. Right panel: Likelihood contours (1,
  2 and 3 $\sigma$) vs. the isotropic fraction and the smoothing angular scale $\sigma$.}
\label{swift}
\end{figure}

Alternative studies with different catalogs were also performed. For
instance, figure \ref{swift}  (left panel) displays the same events
and the distribution of nearby (within 200~Mpc) AGN observed in
X-rays by the SWIFT satellite. The size of the stars in the plot
 is proportional to the measured X-ray fluxes, to a weight
 proportional to the attenuation expected due to the GZK effect and to the
 relative exposure of the observatory in that direction.
 Smoothing out the sources in this map with gaussian windows of a
 given angular scale, and adding a certain fraction $f_{iso}$ of
 isotropic background, a
 likelihood test leads to optimal parameters (displayed in the right
 panel) corresponding to angular scales below $\sim 10^\circ$ and
 isotropic fractions between 40 and 80\%. It is clear that a model
 consisting of only the sources in the catalog with deflections of a
 few degrees would not be consistent with the data. The isotropic
 fraction that is required could well be accounting for 
the  faint or faraway  sources not included in the catalog,
 or for the contribution from a strongly deflected heavy cosmic ray component.
We note that the actual sources of cosmic rays may be different than the 
AGN (e.g. they could be gamma ray bursts, galaxy clusters 
 or colliding galaxies), and hence  in the studies
described the AGN may just be acting as a tracer  of a different but similarly
 distributed population. Also the angular scales inferred are only
 indicative and may not reflect the actual deflection suffered by
 cosmic rays, since the closest AGN to an event need not be its
 source.

It is important that the correlation with nearby extragalactic objects
observed is consistent with 
cosmic rays from more distant sources having lost energy in accordance
with the flux suppression seen in the energy spectrum, and hence this
further supports the interpretation that this suppression is 
 related to the GZK
effect and not just due to the exhaustion of the sources.

\begin{figure}
\centerline{\includegraphics[angle=0,width=0.55\linewidth]{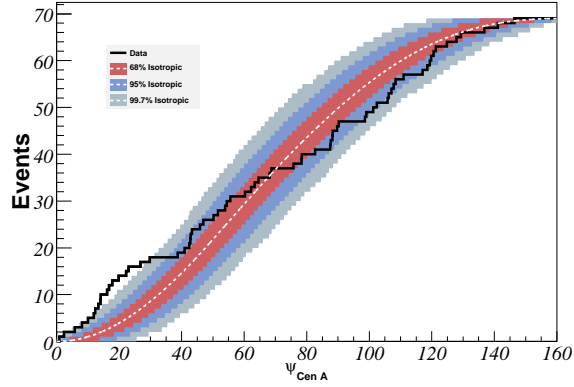}}
\caption{Cumulative number of events vs. the angular distance from 
Centaurus~A, compared to the isotropic expectation. }
\label{cena}
\end{figure}

A significant concentration  of events is found  around
the location of Centaurus~A (corresponding to the largest star in the
left panel of fig.~\ref{swift}), which is particularly interesting
because this AGN lies at only $\sim 4$~Mpc from us. Figure~\ref{cena} shows
the number of observed events as a function of the angular distance 
 from Cen~A  together with the isotropic expectations (average and 68, 95 and
 99.7\% expected isotropic dispersion). 
The most significant departure from isotropy
 is seen for $18^\circ$, for which 13 events are observed while only 3
 are expected. Whether these events come from Cen~A or from other
 sources, such as from the Centaurus cluster lying behind (at $\sim
 45$~Mpc) is still unclear, but this is certainly a region that looks
 specially promising for future anisotropy searches.

\section{Composition}

The other important piece of information one would like to know about
the cosmic rays is their composition, i.e. whether they are protons or
heavier nuclei and if there are some detectable fluxes of 
photons or neutrinos. This knowledge could also help  to better understand the
origin of the different features in the spectrum and the properties of
the acceleration and propagation processes. 

Purely electromagnetic showers, like those initiated by photons,
develop by a combination of $e^{\pm}$ pair production processes by
photons and of electron (or positron) bremsstrahlung, so that after each
interaction length the number of particles in the shower essentially
doubles. Hence, the total number of
particles grows exponentially with the grammage traversed, until the
energies of the individual particles fall below a critical value
$E_c\simeq 86$~MeV for which the $e^\pm $ energy losses by ionization
become important and the shower begins to attenuate. At the
maximum of the shower the number of particles is then $N_{max}\simeq
E/E_c\simeq 10^{11}E/{\rm EeV}$ and the depth of shower maximum
$X_{max}$ depends logarithmically on the energy of the primary (note that the
radiation length in air is $X_0=37$~g/cm$^2$, and the interaction length is
$\lambda\simeq  X_0 {\rm ln}2$, so that there are typically 30-40 interaction lengths
before the ultra-high energy photon shower reaches the maximum).

\begin{figure}
\centerline{\includegraphics[angle=0,width=0.95\linewidth]{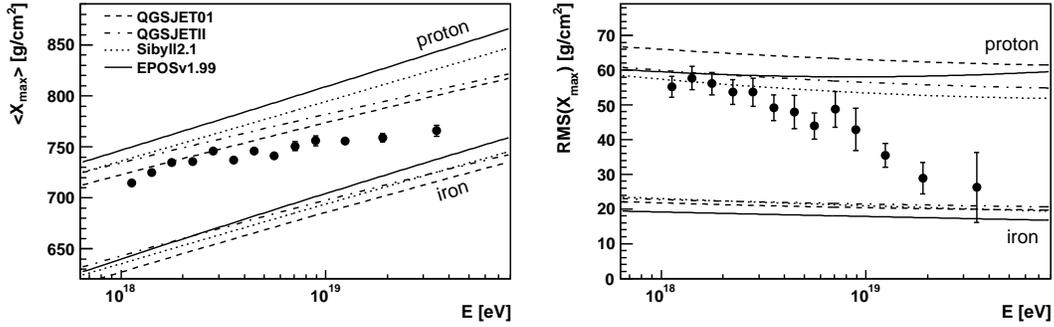}}
\caption{Measured values of $X_{max}$ (left) and its RMS  (right) as a
  function of energy.}
\label{composition}
\end{figure}

Hadronic showers develop differently because in the interaction of a proton
with a nucleus in the air a very large number, of order $\sim 10^2$ at the
highest energies, of
secondaries   are produced.
These secondaries are mostly pions, and the neutral ones (amounting to about
1/3 of the total) immediately decay into
two photons and feed the electromagnetic component of the shower, while the
charged ones will reinteract hadronically producing again a large number of
pions. This process repeats typically for $n\simeq 5$ or 6 times until the
individual pion energies are below a few tens of GeV and the charged pions are
able to decay before reinteracting, producing in this way muons and
neutrinos, which carry away a fraction $(2/3)^n\simeq 10$\% of the primary
energy. The remaining $\sim 90$\% is what went into the electromagnetic
component through the neutral pions of the different generations. Since the
multiplication of particles in an hadronic shower proceeds at a faster rate
 than in the case of photon primaries,
the maximum of the shower is reached earlier. 
The other distinguishing signature of hadronic showers 
is the presence of a sizeable
number of muons reaching the ground.  In the case of primary nuclei, a simple
description can be obtained  with the so-called superposition model, which
considers the shower produced by 
a nucleus of  mass number $A$ and energy $E$ as being
a collection  of $A$ showers produced by nucleons of energy $E/A$. The
resulting shower will then develop earlier (since the depth of maximum of a
proton shower scales as log$E$) and will also have smaller fluctuations, since
the individual maxima  of the $A$ subshowers get averaged out. These two
observables ($X_{max}$ and its fluctuations)  allow then 
to get information  on the cosmic ray composition. The results of the
measurements performed with the Auger fluorescence detectors \cite{augercomp}
 are displayed in
figure~\ref{composition}, together with the predictions for proton and Fe
primaries  obtained using different hadronic interaction models, which
actually need to be extrapolated to energies  well beyond those measured at
accelerators, and are hence still affected by significant uncertainties.  A
transition from a light composition at few EeV towards one approaching  the
expectations from heavier nuclei (even close to those of iron) at $\sim 40$~EeV
is observed. One has to keep in mind that an increase in the proton
nucleus cross section beyond what is considered in the usually adopted
hadronic models would also affect  the inferred nuclear masses since in this
case protons could mimic the expected behavior of heavier nuclei.

\begin{figure}
\centerline{\includegraphics[angle=0,width=0.5\linewidth]{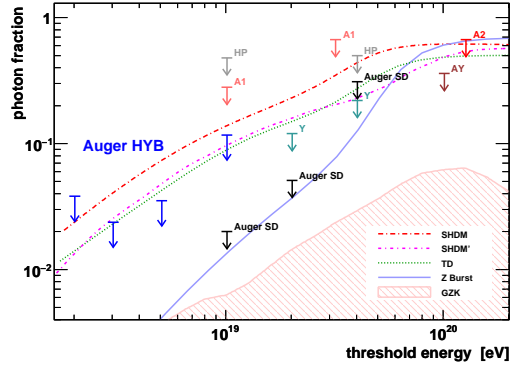}}
\caption{Bounds on the allowed fraction of photons vs. the energy threshold. }
\label{photons}
\end{figure}

The values of $X_{max}$ shown in figure \ref{composition} also indicate that
the fraction of showers that could be produced by photons is small, since
those would be deeply
penetrating and hence lead to $X_{max}$ values larger than the predictions for
protons.  Moreover, a more restrictive constraint on the photon
fraction  can be obtained using the larger statistics of the surface detector
and exploiting the fact that purely electromagnetic showers, having no muonic
component, lead to slower rise-times of the signals in the water Cherenkov
detectors, and also developing deeper in the atmosphere they lead to shower
fronts with smaller radius of curvature. The results of these two measurements
\cite{photonbounds} 
allow then to set the bounds on the photon fraction  displayed in
figure~\ref{photons}, which for instance exclude photon 
fractions larger than 2\% for
$E>10$~EeV. These bounds already exclude many `top-down' models for the
production of ultra-high energy cosmic rays through decays of super heavy
particles or topological defects, since these would lead to significant photon
fluxes (some predictions are shown in the figure), and hence  the standard
scenarios of `bottom-up' acceleration in astrophysical sources is
reinforced. The present sensitivity is still insufficient to detect the
photons that could be produced if ultra-high energy cosmic rays are
extragalactic protons that get attenuated by photo-pion interactions with the
CMB. The neutral pion decays would lead to photon fluxes somewhere in the
shaded region of the plot, which will start to be tested in a few years with
the continuous operation of the Auger Observatory.

\begin{figure}
\centerline{\includegraphics[angle=0,width=0.55\linewidth]{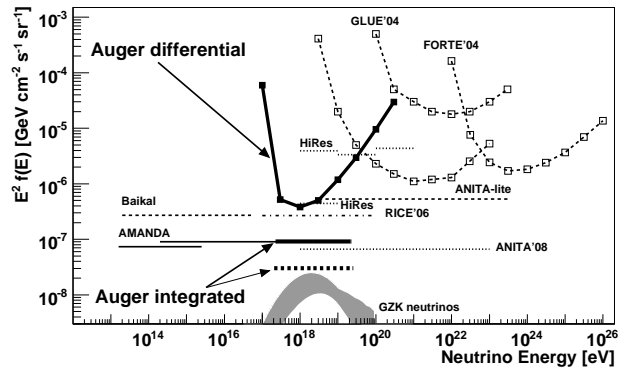}}
\caption{Neutrino bounds }
\label{neutrinos}
\end{figure}

Another important search is that for diffuse neutrino fluxes
\cite{neutrinobounds},  such as those
expected to result from  the decays of the charged pions produced in
the attenuation of extragalactic protons (cosmogenic neutrinos).  
Being weakly interacting, the neutrinos arriving near the vertical have a
small chance to interact in the atmosphere, but on the other hand, neutrinos
arriving near the horizon may have a first interaction not far from the
detector, and hence produce horizontal showers that are young (i.e. with
significant electromagnetic component), unlike the horizontal showers produced
by hadrons that start very far away at the top of the atmosphere and hence
have their electromagnetic component completely attenuated and lead only 
to very
narrow pulses in the detectors due to the surviving muon component.
Another effective way to observe neutrino induced showers is by looking at
 those produced
by tau neutrinos coming from slightly below the horizon. In this case, a charged
current interaction in the rock produces a tau lepton that can travel
several km and eventually exit the ground and decay above the detector,
producing an observable shower. Since neutrino oscillations are expected to
lead to an equal admixture of the different flavors, even in the case that the
sources produce only muon and electron neutrinos by pion decays, tau neutrino
fluxes are also expected. These searches for  upgoing
showers represent actually the most sensitive way to look for diffuse
neutrinos with the Auger Observatory. The resulting bounds are displayed in
figure~\ref{neutrinos}. They are particularly  sensitive at EeV energies,
which is just the energy range were cosmogenic neutrinos are expected to
peak. However, the present sensitivity is still above the
most optimistic predictions (shaded region in the plot) but some improvements
are expected to be obtained with increased statistics. Observation of these
diffuse neutrino fluxes would strongly favor a proton composition at the
highest energies, because heavy nuclei  lead to much smaller expected neutrino
fluxes since having smaller speeds they are below
the threshold for pion production until much higher energies.

\end{document}